\documentclass[prl,floatfix,twocolumn,superscriptaddress,showpacs]{revtex4}

\usepackage{graphicx}

\usepackage{amsmath}

\usepackage{units,xspace}
\usepackage{color}

\newcommand{\Bc}{B_{\rm col}}
\newcommand{\Bp}{B_{\rm p}}
\newcommand{\Bs}{B_{\rm s}}

\newcommand{\dtheta}{\Delta\theta}
\newcommand{\eg}{{e.g., }}

\newcommand{\kB}{k_{\rm B}}
\newcommand{\Pec}{{\rm Pe}}
\newcommand{\Rey}{{\rm Re}}
\newcommand{\tenG}{{\bf G}}
\newcommand{\vecr}{{\bf r}}

\definecolor{rev}{rgb}{0,0.45,0.08}
\definecolor{bl}{rgb}{0,0,1.00}
\begin{document}

\title{Hydrodynamic Pair Attractions Between Driven Colloidal Particles}

\author{Yulia Sokolov}

\author{Derek Frydel}
\affiliation{Raymond and Beverly Sackler School of Chemistry, Tel Aviv
  University, Tel Aviv 69978, Israel}

\author{David G. Grier}
\affiliation{Department of Physics and Center for Soft Matter
  Research, New York University, New York, NY 10003}

\author{Haim Diamant}

\author{Yael Roichman}
\email{roichman@tau.ac.il}
\affiliation{Raymond and Beverly Sackler School of Chemistry, Tel Aviv
  University, Tel Aviv 69978, Israel}

\date{\today}

\begin{abstract}
  Colloidal spheres driven through water along a circular path by an
  optical ring trap display unexpected dynamical correlations.  We use
  Stokesian Dynamics simulations and a simple analytical model to
  demonstrate that the path's curvature breaks the symmetry of the
  two-body hydrodynamic interaction, resulting in particle pairing.
  The influence of this effective nonequilibrium attraction diminishes
  as either the temperature or the stiffness of the radial confinement
  increases.  We find a well defined set of dynamically paired states
  whose stability relies on hydrodynamic coupling in curving
  trajectories.
\end{abstract}

\pacs{82.70.Dd, 87.80.Cc}

\maketitle

Position correlations among particles in thermal equilibrium stem
solely from their potential energy.  Although the motions of
fluid-suspended particles are correlated via flows that they induce
(hydrodynamic interactions), this has no effect on the equilibrium
configurations.  For particles driven out of equilibrium, however,
this is no longer the case, and they may get closer or drift apart in
the absence of an interaction potential.
  Examples are found in studies of nonequilibrium particle aggregation
  and droplet coalescence under flow (\eg \cite{BradyLeal}),
  aggregation of blood cells in channel flow \cite{Fung}, and strong
  correlations in sedimenting suspensions \cite{Ramaswamy}.  Nonequilibrium pair attractions have been reported in
  simulations of particles, embedded in a flowing bath of
  smaller particles \cite{Dzubiella}, or driven through such a bath
  \cite{Oshanin}.  Yet, all these phenomena are observed
  in the presence of either many-body effects or nonuniform flow,
  where spatially separated particles experience a different drive.

It is not self-evident that \emph{two} identical particles, having no
intrinsic interactions, can attract or repel each other under
conditions of \emph{identical} driving \cite{Squires}.  This is
because a uniaxial driving force, breaking rotational symmetry,
is generally insufficient to break the symmetry of the hydrodynamic
interaction on the two-body level.  A familiar example is
an isolated pair of noninteracting colloidal particles, sedimenting
under gravity in an unbounded viscous fluid.  One particle drags the
other to the exact same extent that the other pushes it; hence, their
separation remains constant.  One way to break this symmetry is to introduce a boundary.
Thus, those two sedimenting
particles, when falling away from a horizontal wall, experience a
hydrodynamic force that draws them together \cite{Squires}.  Another
way is to consider more than two particles.  For example, driving a
one-dimensional array of particles by a uniform flow through a slit
creates hydrodynamic restoring forces, resulting in
density waves \cite{Beatus}.

In this Letter we show that the symmetry of the hydrodynamic
interactions of identically driven particles in a fluid can
be  spontaneously broken on the two-body level if their
trajectories curve.  The result is the formation of dynamically paired
states, whose stability determines the steady transport properties of
the system.

\begin{figure}[tbh]
  \centering
  \includegraphics[width=\columnwidth]{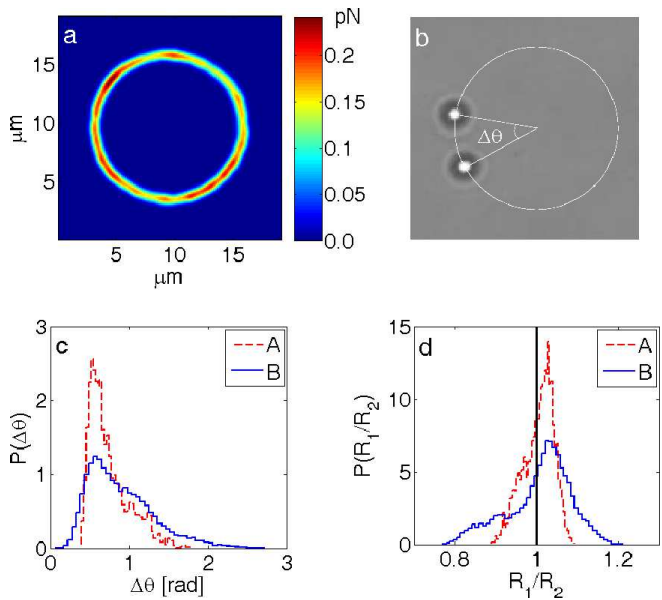}
\caption{(color online)
  (a) Magnitude of the tangential component of the force exerted by an optical vortex (system A) on a single colloidal sphere. (b) Visualization of a pair having an angular
  separation $\dtheta$. (c,d) Distributions of experimentally measured
  angular separation (c) and ratio between the radii of the leading
  and trailing particles (d) in a two-particle vortex for the two
  systems.}
\label{fig:experiment}
\end{figure}

Our samples consist of aqueous dispersions of colloidal polystyrene
particles, sealed in a $50~\unit{\mu m}$ thick gap between a glass
microscope slide and a coverslip.
The Debye-H\"uckel screening length
is of order 10~\unit{nm}, implying that the charge-stabilized spheres'
electrostatic interactions are negligible at micrometer-scale separations.
Individual spheres are captured and driven through water by
extended optical traps, created with the holographic optical trapping
 technique \cite{DufresneCurtis,Polin05}.
In our implementation \cite{Roichman07}, we use a helical mode of light,
which focuses to a ring-like optical trap known as an optical vortex
\cite{DufresneCurtis,Roichman07,Simpson97}.
Particles are drawn to the optical vortex by forces directed by
intensity gradients \cite{Ashkin},
and are propelled around it by radiation pressure arising from the
helical beam's phase gradients \cite{Roichman08}.
Adaptive optics methods are employed to reduce variations in the
tangential driving force \cite{Polin05,Roichman05}.
The optical force profile, as inferred
from the motion of a single particle, is shown in
Fig.~\ref{fig:experiment}(a), revealing a roughness of about 30\%.
The optical vortex is powered with 2.5~\unit{W} of laser light at a vacuum wavelength of 532 nm,
which suffices to drive trapped particles with
tangential velocities $V_\theta$ of a few tens of micrometers per second,
corresponding to
angular velocities $\Omega$ of a few radians per second.
Standard methods of digital video microscopy \cite{Crocker96} are
used to track the particles at a rate of 30~\unit{frames/s}
with 20~\unit{nm} accuracy.

We present results for two sets of experimental parameters (particle
diameter $\sigma$, vortex radius $R$, tangential driving force
$F_\theta$, and radial trapping stiffness $k_r$), as listed in
Table~\ref{tab_parameters}.  Experiments were performed for numbers of
particles ranging from $N = 1$ to $N = 12$, running around the ring in
single-file trajectories.
The particles' inertia may be ignored because of the negligible Reynolds
number, $\Rey = \rho V_\theta \sigma/\eta \sim
10^{-5}$, where $\eta$ and $\rho$ are the viscosity and mass density
of water, respectively.
Random thermal forces contribute fluctuations to the circulating
particles' velocities \cite{ft_thermal}.  The strength of the driving
relative to thermal fluctuations is characterized by the P\'eclet
number, $\Pec = F_\theta\sigma/(\kB T)\simeq 90$ and $74$,
respectively, for systems A and B, where $\kB$ is Boltzmann's
constant, and $T$ the absolute temperature.
The ratio of the radial
confinement strength and tangential driving, $\kappa=k_r R/F_\theta$,
serves as another characteristic parameter.  We estimate $\kappa \simeq
30$ for both systems.

\begin{table}[!b]
\begin{ruledtabular}
\begin{tabular}{ccccc}
system & $\sigma$ ($\mu$m) & $R$ ($\mu$m) & $F_\theta$ (pN) & $k_r$ (pN/$\mu$m) \\
\hline
A & $1.48\pm0.08$ & $6.25\pm0.05$ & $0.25\pm0.08$ & $1.2\pm0.3$ \\
B & $0.97\pm0.03$ & $6.19\pm0.05$ & $0.31\pm0.09$ & $1.7\pm0.8$
\end{tabular}
\end{ruledtabular}
\caption{The two experimental parameter sets: particle diameter
$\sigma$, vortex radius $R$, tangential driving force $F_\theta$, radial confinement
stiffness $k_r$. We calculate $F_\theta$ from the mean tangential velocity in a
single-particle vortex, $F_\theta=V_\theta/\Bs$, and $k_r$ from the single particle's
mean-square radial fluctuations assuming thermal equipartition
\cite{ft_thermal}, $k_r=\kB T/\langle(\Delta r)^2\rangle$.}
\label{tab_parameters}
\end{table}

Because the particles have no intrinsic long-range interactions and
because they are driven by the same optical force, a pair of particles
might be expected to diffuse independently and thus sample all
possible separations with uniform probability.  Experimentally,
however, the two particles enter into a long-lived paired state
\cite{suppl}, as quantified in the histograms of angular separations,
Fig.~\ref{fig:experiment}(c).  The narrower distribution observed for
system A is consistent with its larger $\Pec$.

The observed correlation between the particles' motions must arise
from the interplay of optical and hydrodynamic forces acting on them.
A single sphere circulates around a smooth optical vortex with a
uniform tangential velocity, $V_\theta = \Bs F_\theta$, where $\Bs =
(3\pi\eta\sigma)^{-1}$ is the sphere's Stokes self-mobility.  As a
sphere moves, it creates a flow field $\tenG(\vecr){\bf F_{\theta}}$
at position $\vecr$.  A second sphere placed on the optical vortex at
$\vecr$ experiences not only the optical force $F_\theta$, but also
the drag force due to its neighbor's flow.  To leading order in
$\sigma/r$, the change in the tangential velocity of the second
particle due to its neighbor is simply the local tangential flow
velocity, and $\tenG$ is given by the Oseen tensor
\cite{HappelBrenner}, $G_{ij}(\vecr)=(8\pi\eta r)^{-1} \, (\delta_{ij}
+ r_jr_j/r^2)$, $i,j = x,y,z$.  Applying this result to our case of
two spheres on a ring of radius $R$, separated by an angle
$\theta_{ij} = \theta_j - \theta_i$, we get for the tangential
velocity,
\begin{align}
  V_\theta(\theta_{ij})
  & \simeq
  [\Bs + g(\theta_{ij})] F_\theta, \quad \text{where} \nonumber\\
  g(\theta_{ij})
  & =
  \frac{1 + 3 \cos\theta_{ij}}{16 \pi \eta R \sqrt{2 (1 - \cos\theta_{ij})}}.
  \label{eq:vtheta}
\end{align}
Equation~(\ref{eq:vtheta}) is symmetric under particle exchange,
$\theta_{ij} \rightarrow \theta_{ji} = -\theta_{ij}$, implying that
the tangential component of the hydrodynamic coupling cannot account
for the observed pairing. Because the circulating particles follow a
curved path, however, the flow field due to particle $i$ generally has
a component normal to the trajectory at the position of particle $j$.
(See \cite{suppl} for a theoretical visualization of the
  flow field.) Since the particle is held radially by the trap, this
normal flow component gives rise to a radial force,
\begin{align}
   F_r(\theta_{ij})
   &\simeq
   \Bs^{-1} h(\theta_{ij}) \, F_\theta, \quad \text{where}\nonumber\\
   h(\theta_{ij})
   &= \frac{3\sin\theta_{ij}}{16\pi\eta R \sqrt{2(1 - \cos\theta_{ij})}}.
\label{eq:vr}
\end{align}
This hydrodynamic force displaces the particle radially until it is
counter-balanced by an equal optical force.

\begin{figure}[tbh]
  \centering
  \includegraphics[width=1.0\columnwidth]{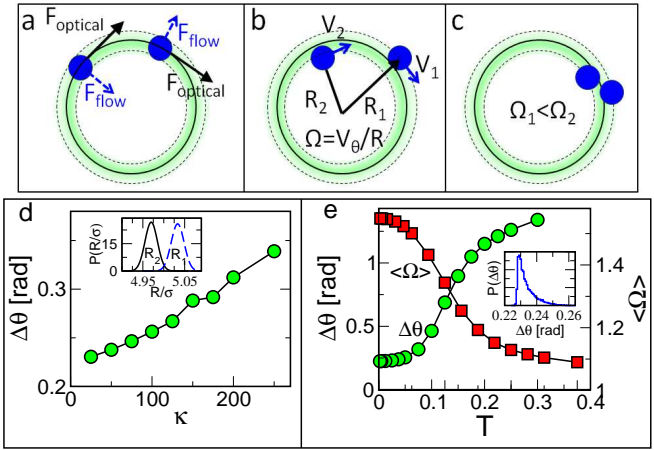}
  \caption{(color online). (a--c) The pairing mechanism.
    As optical forces drive the particles along the ring, the
    resulting fluid flow pushes the leading particle outward and pulls
    the trailing one inward.  Once displaced radially, the particles'
    angular velocities change, making the trailing particle catch up
    with the leading one. (d,e) Simulation results for the effect of
    radial confinement strength (d) and temperature (e) on the average
    angular separation (green circles) and average angular velocity
    (red squares) in a two-particle vortex.  The insets show the
    distributions of radial positions of the leading ($R_1$) and
    trailing ($R_2$) particles, and the pair's angular separation, for $T=0.0375$ and $\kappa=25$.}
  \label{fig:pairing}
\end{figure}

The fact that $h(\theta_{ij})$ is antisymmetric under particle
exchange implies that the two particles are displaced
\emph{oppositely} in the radial direction.  As is depicted
schematically in Fig.~\ref{fig:pairing}, the leading particle is
shifted to a slightly larger radius, $R_1>R$, and the trailing
particle to a slightly smaller one, $R_2<R$.  This
spontaneous symmetry breaking is observed in the
statistics of measured trajectories, plotted in
Fig.~\ref{fig:experiment}(d).  If the tangential force $F_\theta$ is
assumed constant, the radial displacement increases the trailing
particle's angular velocity relative to that of the leading particle,
and so causes the particles to form a hydrodynamically bound pair.

\begin{figure}[tbh]
 \centering
 \includegraphics[width=0.9\columnwidth]{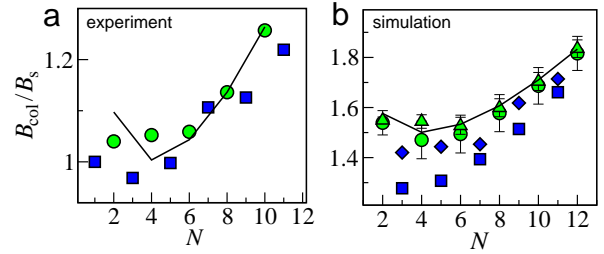}
  \caption{(color online).
    Collective mobility, scaled by the single-particle mobility, as a
    function of particle number. (a) Experimental results for even
     $N$ (green circles) and odd $N$ (blue squares), measured from the
    ratio of mean velocities for $N$ particles and for a single
    particle in system A. (b) Simulation results for even $N$ (green circles, $T=0.1$; green triangles $T=0.02$) and odd $N$ (blue squares, $T=0.1$; blue diamonds, $T=0.02$), at $\kappa=25$.
    The solid lines are theoretical fits for even $N$ [Eq.\ (\ref{collective})] with
    $\alpha=1.10\pm0.02$ (experiment) and $\alpha=1.57\pm0.05$ (simulation).
    Error bars are smaller than symbols unless otherwise indicated.}
\label{fig:collective}
\end{figure}

Pairing also appears in the dynamics of systems comprising more than two spheres.  Driven rings of hydrodynamically coupled spheres thus constitute an interesting model system for studying spontaneous symmetry breaking in many-body systems out of equilibrium.
Figure~\ref{fig:collective}(a) shows
the measured collective mobility as a function of particle number $N$.
The increase in the circulation rate due to a decrease in the
hydrodynamic drag is more pronounced for even values
of $N$.
In systems with odd $N$ the tendency to form pairs leaves one
particle unpaired, which reduces the influence on the
collective drag.

Let us focus on the case of even $N$ and consider an idealized
configuration of $N/2$ pairs, evenly positioned along the ring.
In the
absence of hydrodynamic interactions, each pair will move with
tangential velocity $V_\theta = \Bp F_\theta$, where $\Bp = \alpha\Bs$
is the mobility of a pair ($\alpha$ being a prefactor of order $1$).
Including tangential hydrodynamic couplings as above, we get for the
collective mobility,
\begin{equation}
  \frac{\Bc}{\Bs} = \frac{V_\theta(N)}{V_\theta(1)}
  = \alpha + 2\Bs^{-1} \sum_{n=1}^{N/2-1}
  g(4\pi n/N),
  \label{collective}
\end{equation}
where $g(\theta)$ has been defined in Eq.~(\ref{eq:vtheta}).
In Fig.~\ref{fig:collective}(a) we compare the experimentally measured
collective mobility, for even $N$, with the one predicted by Eq.~(\ref{collective}),
using the vertical shift $\alpha$ as a fitting parameter.
Despite the simplifying assumptions underlying Eq.~(\ref{collective}),
a qualitative correspondence is obtained for $\alpha = 1.10 \pm 0.02$.
We attribute the larger discrepancies for $N = 2$ and $N = 4$ to larger
deviations from the idealized configuration of fixed pairs in these
dilute systems.



We systematically investigated how the pairing dynamics depend on
$\Pec$ and $\kappa$ by performing Stokesian Dynamics simulations
\cite{Brady88}.  The simulated particles are restricted to the
plane and confined to a ring of
radius $R = 5\sigma$ by a harmonic radial potential of stiffness
$k_r$.  A constant tangential force $F_\theta$ drives the particles
around the ring, and a repulsive Weeks-Chandler-Anderson pair
interaction \cite{Weeks71} accounts for collisions.  Hydrodynamic
interactions between particles are modeled using the many-body
Rotne-Prager mobility tensor
\cite{RotnePrager,ft_Rotne}. Temperature is introduced
via a Gaussian-distributed random force that obeys the
fluctuation-dissipation relation using the same mobility tensor.  The
angular velocity is measured in units of $\Bs F_\theta/R$ and the
simulation temperature is measured in units of $F_\theta\sigma^2/(\kB
R)$.  In these units, the P\'eclet number is $\Pec = (R/\sigma)T^{-1}
= 5/T$.

Figure~\ref{fig:pairing}(d) presents simulation results for two
particles on a ring, confirming the role of radial symmetry breaking
in creating particle pairs.  At low temperature and weak radial
confinement, the particles form permanent pairs. Paired particles
follow trajectories of different radii  with $R_1/R_2 \simeq 1.01$ [Fig.~\ref{fig:pairing}(d)
inset]. Pairing in simulation is more pronounced than in experiment, with a smaller and more narrowly distributed angular separation [Fig.~\ref{fig:pairing}(e)
inset] \cite{ft_approach}. For infinitely strong
radial confinement, no pairing is observed.  Vigorous diffusion at
elevated temperatures overcomes the hydrodynamic pairing mechanism
and allows pairs to move apart.  This increases their drag, and
reduces their mean angular velocity [Fig.~\ref{fig:pairing}(e)].

Simulations for rings with even $N>2$ exhibit pairing as
well.  Comparing the collective mobilities with our simplified model
[Fig.~\ref{fig:collective}(b)], we obtain a reasonable fit using the
value $\alpha = 1.57 \pm 0.05$.  The value of $\alpha = \Bp/\Bs$ can
be related also to the inter-particle distance through the theoretical
expression for the mobility of a pair of spheres
\cite{HappelBrenner}.  In the simulation we measure an
angular separation of $\dtheta = 12.9 \pm 0.1^\circ$, which yields
$\alpha = 1.52 \pm 0.02$.  These two independent measurements of the
pair mobility agree to within the statistical error.  The smaller
experimental value of $\alpha$ [Fig.~\ref{fig:collective}(a)] suggests
still weaker pairing.
These quantitative differences between
experiment and simulation may be related to such effects as the roughness of the
experimental force landscape \cite{ft_thermal} and a possible nonintrinsic repulsion \cite{ft_approach}.

At low temperatures, particles may assemble into different
configurations as a result of their tendency to form pairs, giving rise to a rich
behavior of degenerate limit cycles.  The simplest example is the case
of three particles \cite{Roichman07,ReichertStark}.  The breaking of a
triplet into a faster pair and a slower single particle exists also in
the absence of curvature and radial freedom, yet, in simulations we
observe long-term pairing and the consequent limit cycle of an
alternating pair only when the radial confinement is made sufficiently
weak.

\begin{figure}[tbh]
\centering
\includegraphics[width=0.7\columnwidth]{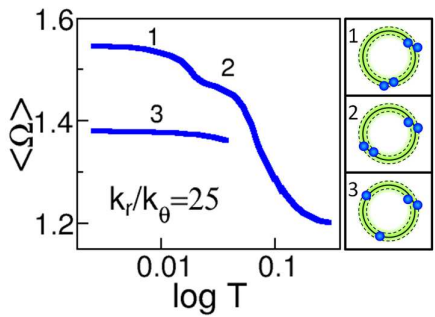}
  \caption{Simulation results for the normalized average angular velocity
    $\Omega$ and particle configuration, as a function of temperature,
    in a four-particle vortex. Insets illustrate the three
    configurations.}
  \label{fig:sim_4p}
\end{figure}

A simulated four-particle vortex is found to converge into three
stable limit cycles, shown in Fig.~\ref{fig:sim_4p}.  For a given
radial confinement strength, particles adopt configurations 1 and 3 at
low temperatures and configurations 2 and 3 at higher temperatures.
With increasing temperature the pairs gradually break.
In configuration 1
the two pairs are separated by a certain angle ($0.6\pi$ in our
simulation), which seems unrelated to the system's geometry.
The transition between configurations 1 and 2 explains the difference in
collective mobility seen in Fig.~\ref{fig:collective}(b). Since
configuration 2 is compatible with our calculation of equidistant
pairs, we get a better agreement for the higher temperature in which
it is stabilized.
This effect might also contribute to the increased
mobility of four particles seen experimentally
[Fig.~\ref{fig:collective}(a)].
The assumed experimental P\'eclet numbers \cite{ft_thermal}
correspond to simulation temperatures of $T = 0.055$ and $0.068$ for
systems A and B, respectively, with $\kappa$ values around $30$. This might indicate that
the experiments lie in the intermediate region between pairing
and unpairing (Fig.~\ref{fig:sim_4p}).
As the number of particles increases so does the number of limit cycles,
leading to a more complex picture to be presented elsewhere.

The spontaneous symmetry breaking and dynamic coupling presented here can arise whenever particles are driven
through a fluid medium along curving trajectories, provided that they
are allowed to slightly move in the radial direction.  Such
nonequilibrium pair attractions, therefore, should affect the behavior
of diverse out-of-equilibrium colloidal systems. In addition, we
intend to use this model system to study further issues of
nonequilibrium physics in a well controlled environment.

This study was supported by the Israel Science Foundation (Grants Numbers 1271/08, 588/06, and 8/10).


\begin{thebibliography}{13}
\bibitem{ft_thermal}
  The issue of whether thermal fluctuations or the vortex
  inhomogeneities are the dominant source of noise remains unclear and
  requires a more detailed study.
\bibitem{ft_Rotne}
This description is strictly valid for sufficiently large
inter-particle distances. Yet, since the pairs have a finite
separation \cite{ft_approach}, we do not expect the deviations at
short distances to lead to a qualitatively different behavior.
\bibitem{ft_approach}
  The minimum interparticle distance within a pair is about $1.7$ particle diameters in the experiment and $1.1$ in the
  simulation. We currently cannot account for the nonintrinsic repulsive effect in
  the experiment that might underly this difference.
\end{thebibliography}
\end{document}